\documentclass[fleqn,usenatbib]{mnras}

\usepackage{newtxtext,newtxmath}
\newcommand{\xmark}{\ding{55}}

\usepackage[T1]{fontenc}

\DeclareRobustCommand{\VAN}[3]{#2}
\let\VANthebibliography\thebibliography
\def\thebibliography{\DeclareRobustCommand{\VAN}[3]{##3}\VANthebibliography}

\usepackage{graphicx}	
\usepackage{amsmath}	
\usepackage{float}
\usepackage{pifont}	

\usepackage[dvipsnames]{xcolor}
\usepackage{tikz}

\definecolor{lime}{HTML}{A6CE39}
\DeclareRobustCommand{\orcidicon}{
	\begin{tikzpicture}
	\draw[lime, fill=lime] (0,0) 
	circle [radius=0.13] 
	node[white] {{\fontfamily{qag}\selectfont \tiny ID}};
	\draw[white, fill=white] (-0.0625,0.035) 
	circle [radius=0.007];
	\end{tikzpicture}
	\hspace{-2mm}
}

\foreach \x in {A, ..., Z}{\expandafter\xdef\csname orcid\x\endcsname{\noexpand\href{https://orcid.org/\csname orcidauthor\x\endcsname}
			{\noexpand\orcidicon}}
}

\title[DEM L50 models]{A comprehensive hydrodynamical study of SB DEM L50: understanding off-centre SNe and soft X-ray luminosity}

\author[R.~Orozco-Duarte et al.]{Rogelio Orozco-Duarte$^{1\orcidA}$\thanks{E-mail: rorozco@astro.unam.mx}, Guillermo Garc\'{i}a-Segura$^{1\orcidB}$, Aida Wofford$^{1\orcidC}$ and Jes\'{u}s~A.~Toalá$^{2\orcidD}$
\\
$^{1}$Universidad Nacional Autónoma de México, Instituto de Astronomía, AP 106,  Ensenada 22800, BC, México\\
$^{2}$Instituto de Radioastronomía y Astrofísica, UNAM Campus Morelia, Apartado postal 3-72, 58090 Morelia, Michoacán, México.
}

\date{Accepted XXX. Received YYY; in original form ZZZ}

\pubyear{2023}

\begin{document}
\label{firstpage}
\pagerange{\pageref{firstpage}--\pageref{lastpage}}
\maketitle

\begin{abstract}
The superbubbles (SBs) carved in the interstellar medium by stellar winds and supernovae (SNe) are filled with hot ($T>$10$^{6}$~K) gas that produces soft X-ray emission (0.3-2.0\,keV). Models that assume a constant density medium and central SNe events fail to reproduce the soft X-ray luminosity that is observed in some SBs. 
We address this problem by generating models that trace the history of SNe in the SB and that produce off-centre SNe, that account for the missing soft X-ray emission. 
We test the models against archival, radio, optical, and X-ray observations of the SB DEM\,L50 located in the Large Magellanic Cloud. 
The soft X-ray properties of DEM L50, including its high luminosity, makes it a perfect candidate to test our models. Furthermore, the multiple wave-band observations of this object will help us to assess how well our models can reproduce other SB properties beside its soft X-ray properties. We find that a configuration where DEM L50 forms at the edge of a filament reproduces the observed soft X-ray luminosity, optical morphology, shell velocity, and swept-up mass of neutral gas. This configuration is supported by IR observations of the LMC. In addition, we find that off-centre SNe, which enhance soft X-ray emission, naturally occur for all of the initial ambient conditions we tested in our models. Finally, we show that an off-centre SN can explain the observed soft X-ray luminosity of DEM L50, and that the resulting luminosity is consistent with a plasma in non-equilibrium ionisation. 
\end{abstract}




\section{Introduction}
Stellar winds and supernovae (SNe) generated by massive stars in stellar clusters inject energy and momentum to the interstellar medium (ISM). This causes the displacement of the surrounding material, resulting in the formation of low-density, high-temperature cavities called superbubbles (SBs, \citealt{Chu2008}). These structures can have sizes of $\sim$100\,pc \citep{Kavanagh2020} and the shocks produced by the stellar feedback can produce star formation \citep{Oey2005, Krause2018}. As SBs are filled with hot gas with temperatures of the order $T\sim$\,10$^{6}$\,K, they produce diffuse soft X-ray emission. This diffuse emission has been observed in multiple objects \citep{Chu1990, Cooper2004, Jaskot2011, Reyes-Iturbide2014} and can have a very complex structure \citep{Kavanagh2015}.

One key aspect that many research groups have tried to understand, is the evolution of the energy budget in SBs. Specifically, the soft X-ray luminosity of some SBs is higher than the predicted value by the classical pressure-driven model of \cite{Weaver1977} by more than an order of magnitude \citep{Chu1990, Jaskot2011, Rodriguez2011,Reyes-Iturbide2014}. There are different proposed solutions that could fix this discrepancy, including off-centre SNe within the SB \citep{Chu1990}, mass-loading due to ablation of cold clumps \citep{Martin1995}, and high-metal content due to the high stellar mass loss \citep{Silich2001}.

One or a combination of the previous solutions have been included in numerical simulations to tackle this problem \citep{Velazquez2013, Krause2014, Rogers2014, Castellanos-Ramirez2015, Schneiter2022}. 
In particular, SBs with X-ray bright limbs are associated with recent SN activity and high X-ray luminosities; and models with off-centre SNe reconcile models and observations. For example, \citet{Rodriguez2011} and \citet{Schneiter2022} used hydrodynamical simulations that include stellar winds from the current massive star population to model the SB LMC N\,70 (a.k.a. DEM L\,301). 
The problem with this solution that reconciles theory and observations is that there is no explanation for how a SN got off-centre.
Star-formation in SBs seems to be a hierarchical process, on which the youngest populations of stars are primarily located at the edge of SBs  \citep{Oey2005, Krause2018, Barnes2022}.
Thus, old stars transitioning into SNe are not expected near the edge of a SB.

\begin{figure*}
    \centering
    \includegraphics[width=\linewidth]{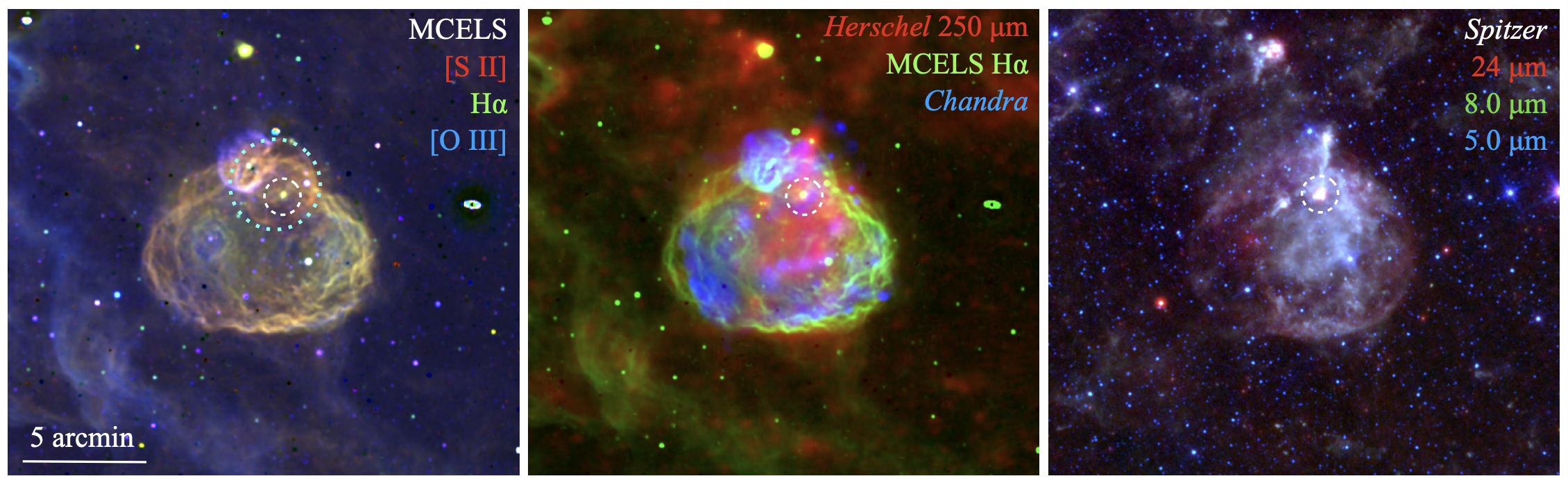}
    \caption{Colour-composite images of DEM L50. {\it Left}: Optical image composed by the continuum-subtracted, narrow-band images obtained from the MCELS \citep[Magellanic Cloud Emission-line Survey][]{Smith2005}. {\it Middle}: IR, optical and X-ray emission from {\it Herschel} 250~$\mu$m, MCELS H$\alpha$ and {\it Chandra}, respectively. {\it Right}: IR image obtained by combining the {\it Spizer} 24, 8.0 and 5.0~$\mu$m images. The position of Gr~3 is shown with a (white) dashed-line circle. The northern blister is shown with a (cyan) dotted-line circle. All panels have the same field of view (FoV). North is up, east to the left.}
    \label{fig:DEML50}
\end{figure*}

Not many authors have modelled SBs forming in more realistic, non-uniform density environment. However, some attempts have been presented in the literature \citep[see for example][]{Rogers2013,Rogers2014}.
In this work, we aim to produce an off-centre star cluster and SNe by varying the properties of the stellar population and environment where the SB forms.
To do this, we produced detailed models of SB DEM L50. We also test different SB-formation scenarios in order to explain the X-ray luminosity of this object.

We selected DEM L50 because it has been previously studied at almost every wavelength. In Figure~\ref{fig:DEML50}, we show the morphology of DEM L50 in different wavebands. In the leftmost panel of Figure~\ref{fig:DEML50} we show the optical shell as imaged by the Magellanic Cloud Emission-line Survey \citep[MCESL;][]{Smith2005}. Projected on the sky, DEM L50 has an elliptical shape with major and minor axes of 9$\arcmin$ (130\,pc) and 7$\arcmin$ (100\,pc) respectively. A couple of cavities can be seen in the northern part of DEM L50. The middle and rightmost panels of Figure~\ref{fig:DEML50} show a combination of publicly available images obtained from different telescopes. These include IR observations from {\it Herschel} and {\it Spitzer} which were retrieved from the NASA/IPAC Infrared Science Archive\footnote{\url{https://irsa.ipac.caltech.edu/frontpage/}}. The middle panel also includes a soft X-ray image (0.2--1.5~keV) obtained by processing the available {\it Chandra} data \citep[see][for details]{Jaskot2011}.

As demonstrated by \citet{Jaskot2011}, Figure~\ref{fig:DEML50} corroborates that the X-ray emission is enclosed by the optical shell with bright patches at the southern regions of this SB. Interestingly, the maximum of the diffuse X-ray emission is anti-correlated with the presence of dust. The {\it Herschel} and {\it Spitzer} images presented in the middle and rightmost panels of this figure show a maximum of IR emission towards the centre of DEM L50. 
Furthermore, we observe that the dust distribution is similar to the hydrogen column density, N(H{\sc I}), reported by \cite{Oey2002}. Here, the N(H{\sc I}) is mostly concentrated at the centre of the SB and covers the majority of the optical shell.



Similarly to other SBs, the observed X-ray luminosity of DEM L50 has been reported to be an order of magnitude larger than that estimated by the pressure-driven model of \cite{Weaver1977}, with $L_\mathrm{X}$= 2.2$\times10^{36}$\,erg\,s$^{-1}$ \citep{Jaskot2011}. In this work, we test our models against all these observables. 
By doing this we intend to find the most likely scenario that gave DEM L50 its observed properties and produced its off-centre SN.
The three models we studied are: i) a bow shock; ii) an ambient density that decreases as a power law, 
and iii) a SB expanding at the edge of a filament.

The structure of this paper is as follows. In Section~\ref{sec:Methods}, we briefly describe the stellar population of DEM L50, the estimated mechanical energy which is subsequently used for the SB models, the set of initial conditions for each simulation, and we compare the photoionised-gas radius and velocity of models in which the number of massive stars is different. 
From these models, we select the one that better fits a series of conditions. 
In Section~\ref{sec:Results}, we first present density maps corresponding to different temperature ranges and compare our results with observations of DEM L50. In addition, we compare the optical radius, shell velocity, and swept-off mass of neutral gas with the reported values of DEM L50.  Finally, we present the predicted X-ray luminosities. In Section~\ref{sec:Discussion}, we discuss evidence that DEM L50 is interacting with a filament. We also discuss the role of thermal conduction in our predictions of the X-ray luminosity. In Section~\ref{sec:Conclusions}, we summarise and conclude.


\section{Methods} 
\label{sec:Methods}

\subsection{Stellar population in DEM L50} \label{subsec:Gr3}

The models presented here use the stellar content inferred from the initial stellar mass function (IMF) \textbf{\cite{Oey1996_dynamics}}. Thus, our results also include the history of SNe of this SB. 
In this study, we assume that stars producing winds and SNe are located at the centre of our grid. As we will see in Section~\ref{sec:Results}, the model that reproduces both the X-ray luminosity and the shape of DEM L50 also predicts a stellar cluster that is in a similar position as Gr3 on the northern side of DEM L50 \citep{Rosado1990}. The position of this stellar cluster is shown in the three panels of Figure~\ref{fig:DEML50} with a (white) dashed-line circle. Gr3 has an angular diameter of 0.5~arcmin \citep{Glatt2010}, it is an H$\alpha$, IR and X-ray source as shown in Figure~\ref{fig:DEML50}.

The presence of H$\alpha$ emission from Gr3 is thought to be due to a small association of early-type stars \citep{Rosado1990, Glatt2010}. The observed X-ray emission indicates the presence of stellar winds, which are caused by the collision of the winds. Lastly, the FIR emission could be dust of older massive stars that went through a Wolf-Rayet (WR) phase before exploding as SNe \citep[][]{Toala2015}. This evidence supports the idea that the stellar population in Gr3 could be the powering source of the SB instead of the stars disperse throughout the SB. Furthermore, in Section~\ref{sec:Results} we show that the age of the SB for all of our models is of the order of 5 Myr. According to studies of star clusters in the LMC, such as the one from \cite{Wilkinson2003}, star clusters typically have a radius $\sim$\,2\,pc during a lifespan of 10$^{7}$--10$^{9}$\,Myr. This means that given the estimated time for the SB to grow to the observed diameter ($\sim$ 5\,Myr according to our simulations), the star cluster has not have enough time to disperse more than 2\,pc thus, the Gr3 scenario is plausible.

While we have previously suggested that Gr3 may be the wind source causing expansion of DEM L50, there is still a lack of concrete evidence to support this theory. To further investigate this possibility, it would be useful to confirm the presence of massive stars in Gr3 through new observations. \cite{Rosado1990} studied the stellar population in Gr3 and identified the presence of an O9\,V star. However, there have been no recent studies of this region that would provide up-to-date information on the spectral classification of stars in Gr3.
The VLT FLAMES survey has not revealed any new data on DEM L50 \citep{Evans2011}. The ULLYSES survey is obtaining new data in DEM L50 but for the massive star SK 70-32, however, this star is not in the Gr3 area (see \url{https://ullyses.stsci.edu/ullyses-targets-lmc.html#ullysessample}). Additional studies of the stellar population in Gr3 are needed in order to fully understand if this stellar population could have produced DEM L50.

\subsection{Mechanical energy due to winds and supernovae}

\begin{figure}
    \centering
    \includegraphics[width=1\columnwidth]{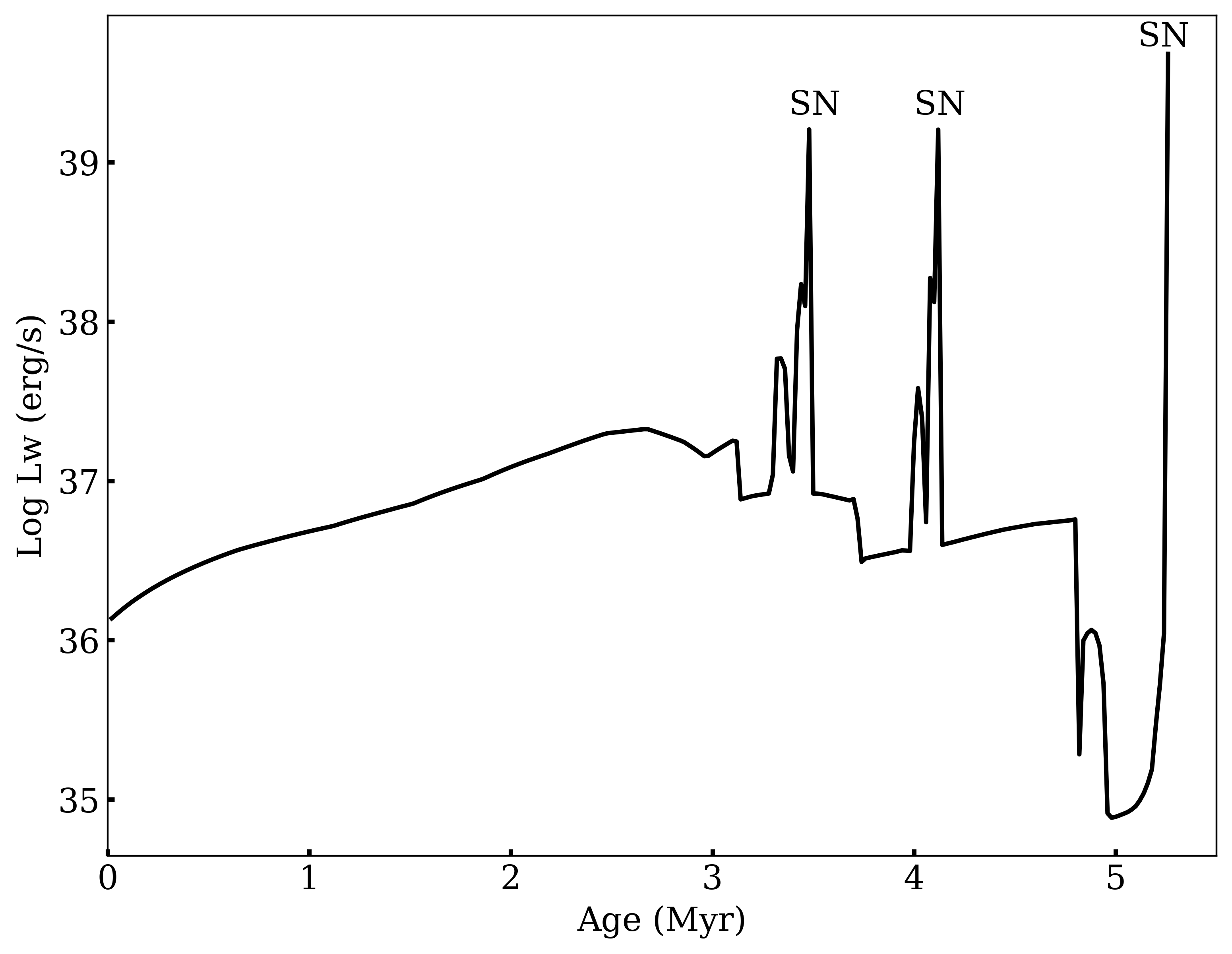}
    \caption{Mechanical wind luminosity as a function of time in units of Myr for model F (see Section~\ref{stars_number} for details). The time at which each star explodes as a SN is labelled on the graph.}
    \label{fig:geneva_track}
\end{figure}

The mechanical energy that a star delivers into the surrounding medium is:
\begin{equation}\label{eqn:1}
    L_w = \frac{1}{2} v_{\infty}^{2}\dot M,
\end{equation}
where $\dot M$ is the mass-loss rate of the central star and $v_{\infty}$ is the terminal wind velocity. This model was developed for a single source that injects a stellar wind at a constant rate. 
Now, to model a star cluster we can just add the contributions of each individual star in the cluster to obtain a total mechanical energy function:
\begin{equation}\label{eqn:2}
    L_\mathrm{tot}(t) = \sum_{i=1}^{N} L_{w,i}(t),
\end{equation}
Here, $L_{w,i}$ is the mechanical wind luminosity of the $i$-th star in a cluster with a total of $N$ massive stars. \cite{Weaver1977} developed the pressure-driven model for a constant mass-loss rate, but as we know from stellar evolution, stars go through different evolutionary stages and this causes the mass-loss rate of a star to be a function of time. 
To capture this behaviour we follow the method described in \cite{Oey1994, Oey1995}, which tracks the changes in the mechanical energy of a star with the aid of a stellar evolution model.
For a detailed discussion of the method we refer the reader to the original work of \cite{Oey1994}, on which the authors define the steps involved to produce the mechanical energy function of a single star.
This method only produces the mechanical energy of the star during its life right before exploding as a SN thus, it does not take into account the energy for this stage. To take into account this contribution, we add the equivalent to the typical energy of a SN at the end of each evolutionary track. We used an energy of 10$^{51}$\,erg, which is the typical energy of type I b/c SN \citep{Janka2012}. 
For this work, we use the stellar evolutionary models of \cite{Schaerer1993}, this models provide tabulated mass-loss values calculated from the empirical law of \cite{deJager1988} and are properly scaled for the metallicity of the LMC. 
Even if there are now more up-to-date mass-loss rates that use modern prescriptions for the mass losses, such as the ones from \cite{Vink2001}, these mass-loss rates are similar to the ones from \cite{deJager1988} \citep[see also][]{Smith2014}. 
Thus, the overall SB properties should be similar integrated in time independently of the mass-loss recipe used to calculate the mechanical energy of the star cluster.
An example of an energy input is shown in Figure~\ref{fig:geneva_track}, were we present the mechanical luminosity function for a star cluster that corresponds to that of model F (see Section~\ref{stars_number} for details). 
The changes in luminosity show the different evolutionary stages of stars. First, the luminosity starts to increase monotonically due to an 85\,M$_{\odot}$ star. 
Then, the following changes between 3 to 3.48\,Myr correspond to this star transitioning to an LBV and then to a WR phase before exploding as a SN (the earliest SN in the Figure). 
The following change in luminosity correspond to the transition from MS to WR of a 60\,M$_{\odot}$ star. Finally, the last change in luminosity is due to three 40\,M$_{\odot}$ stars transitioning from the MS to RSG before exploding as SNe.

\subsection{Superbubble models}
\label{subsec:conditions}

In the present study, we aim to reproduce the X-ray luminosity of DEM L50 by comparing different formation scenarios. The initial conditions for each individual model can be found in Table~\ref{tab:1}. In the following sections, we describe in detail each scenario.

\begin{table}
\begin{center}
\caption{Initial parameters for our three cases of study. Column (1) Shows the name of the scenario. Column (2) shows the initial density of the ISM. Column (3) Shows the initial temperature of the ISM. Column (4) shows the radius of the core measured in pc for the molecular cloud model. Column (5) shows the separation of the stellar cluster of the centre of the molecular cloud in the case of the molecular cloud model. Column (6) shows the velocity of the stars on the bow shock model.}
\begin{tabular}{cccccc}
\hline
Scenario & $n$        & $T_\mathrm{ISM}$ & $r_\mathrm{core}$ & $r_{0}$ & $v$ \\
         & (cm$^{-3}$)& (K)              & (pc)              & (pc)    & km~s$^{-1}$ \\
\hline
Bow shock  & 13       & 100 & \dots & \dots & 10 \\
Cloud      & 10$^{4}$ & 100 & 1.6   & 0.2   & \dots \\
Filament   & 13       & 1   & \dots & \dots & \dots \\
\hline
\end{tabular}
\label{tab:1}
\end{center}
\end{table}


\subsubsection{The star cluster associated with DEM L50 is not at rest} 
Our first scenario supposes that DEM L50 formed in a constant density medium and the cause of the observed elliptical shape is an effect produced by the movement of the star cluster through space, e.g., a bow shock-like structure. In this model, we assume that the velocity of the star cluster is v = 10\,km\,s$^{-1}$.

\subsubsection{The density of the medium is not constant} 
Star clusters are born in a density ambient that is not uniform (or homogeneous). Instead, star clusters are formed in the high density cores of molecular clouds were conditions are proper for star formation. A model for a birth cloud consisting of an internal core radius $r_\mathrm{c}$ and a halo in hydrostatic equilibrium has been explored by \cite{Franco2007}. Following the mathematical treatment of this article they present a series of equations for the density profile and the gravitational acceleration of a cloud with a density distribution $\propto r^{-2}$. This slope in density is what is most typical in clouds \citep{Arthur2006, Arthur2007}. According to equations (6) and (7) of \cite{Franco2007} the density profile of the cloud is:

\begingroup
\large
\begin{equation} \label{eq:2}
    \rho\,(r) = \begin{cases}
    \rho_{0}\;\exp{[-(r/r_\mathrm{c})^{2}]} & \mathrm{for}\;r \leq r_\mathrm{c}\\
    \\
      \dfrac{\rho_{0}}{e}\;\left(\dfrac{r}{r_\mathrm{c}}\right)^{-2} & \mathrm{for}\;r \ge r_\mathrm{c}\\
    \end{cases},
\end{equation}
\endgroup
and the acceleration due to gravity is,
\begingroup
\large
\begin{equation} \label{eq:3}
    g(r) = \begin{cases}
    2c_\mathrm{s}^{2}/r_\mathrm{c}\; \dfrac{r}{r_\mathrm{c}} & \mathrm{for}\; r \leq r_\mathrm{c} \\
    \\
    2c_\mathrm{s}^{2}/r_\mathrm{c}\; \left(\dfrac{r}{r_\mathrm{c}}\right)^{-1} & \mathrm{for}\; r \ge r_\mathrm{c}\\
    \end{cases}
\end{equation}
\endgroup
Here, $\rho_{0}$ is the central cloud density of the cloud and $c_\mathrm{s}$ is the isothermal sound speed. In order to generate a non-spherical distribution we have placed the stars off-centre of the core at a distance $r_{0}$. To find a model that resembles DEM L50, we used realistic conditions found for clouds. We used the parameters reported in \cite{Goldsmith1987} for the core radius and density of the birth cloud and tested various combinations for $r_\mathrm{c}$, $\rho$(r) and the off-centre radius $r_{0}$ until we found the best model.

\subsubsection{The star cluster was born at the edge of a high density filament}

The LMC consists of voids, shells, filaments and a spiral-like galaxy structure as seen in H\,{\sc i} observations \citep{Rohlfs1984, Sungeun1999}. The last scenario we explore is the one of a star cluster that formed at the edge of a filament. This would cause the SB to expand between a two-layer density system. For this case, we set the density of the environment to be the estimated H\,{\sc i} density of DEM L50 \citep[$n\approx$13~cm$^{-3}$;][]{Oey2002}). For the filament, we set the density to $n_\mathrm{filament}$= 260~cm$^{-3}$. This is meant to emulate the density jump due to the shock caused by the filament moving through space. 
This would produce an initial symmetric SB growing inside a constant density ambient that is then depressurised after reaching the edge of the filament, causing an accelerate expansion into the low-density region.


\subsection{Numerical setup}

Our numerical simulations were performed using the {\sc zeus-3d} code  \citep[version 3.4;][]{Stone1992, Clarke1996}. We use spherical coordinates $r$ and $\theta$ to solve the equations of mass, momentum and energy by means of a finite-differences scheme, fully-Eulerian code. We derive the localisation of the ionisation front assuming that ionisation equilibrium holds at all times and that the gas is fully ionised. The location of the ionisation front is given by:

\begin{equation}
    \int n^{2}(r)r^{2}\, dr \approx \dfrac{Q_{0}}{4\pi \alpha_{B}},
\end{equation}
where $\alpha_\mathrm{B}$ is the Case B recombination coefficient and $Q_{0}$ is the stellar H-ionising emission rate. For the current problem we did not implement magnetic fields. All of our simulations are defined in a 400\,$\times$\,400 zones in $r$ and $\theta$, respectively, with an angular extent of 180$^{\circ}$ and a radial extent of 110\,pc. The energy of winds and SNe are injected into a 1.1\,pc spherical zone at the centre of the grid and with a terminal wind velocity of 3000 km\,s$^{-1}$.

\subsection{Mechanical energy input due to massive stars}\label{stars_number}
\cite{Oey2004} presented 1D models for DEM L50 using a mechanical energy input that included massive stars down to 20\,M$_{\odot}$. On the other hand, table 2 in \cite{Oey1996_dynamics} reports that the stellar content in DEM L50 is more than 20 massive stars that range between 85 to 12\,M$_{\odot}$. Note that the maximum observed is 40\,M$_{\odot}$ and the number of more massive stars and SNe is indeed unknown. We see important to test if there are significant differences in the dynamics and evolution of a SB depending on the number of massive stars that are producing the mechanical energy to grow the shell. If there are, including the full stellar population could be important as this could impact the growth of the SB. To investigate this, we carried out 1D simulations to investigate these differences between SBs with different numbers of massive stars. For this purpose, we used different models (A--F) that include different combinations of stars. These are listed in Table~\ref{tab:2}.


\begin{table}
\begin{center}
\caption{Test configurations of SBs in which we vary the number of massive stars. Column (1) shows the model's name and Column (2) shows the different combination of massive stars for each 1D simulation. Column (3) shows the time in Myr at which each massive stars exploded as a SN.}
\begin{tabular}{ccc}
\hline
Model name & Combination of massive stars & $t_\mathrm{SN}$  \\
           &                              & (Myr)            \\ \hline
A          & 85 M$_{\odot}$               & 3.48             \\
B          & 85 + 60 M$_{\odot}$          & 3.48, 4.12       \\
C          & 85 + 40 M$_{\odot}$          & 3.48, 5.26       \\
D          & 85 + 60 + 40 M$_{\odot}$     & 3.48, 4.12, 5.26 \\
E          & 85 + 60 + 2x40 M$_{\odot}$   & 3.48, 4.12, 5.26 \\
F          & 85 + 60 + 3x40 M$_{\odot}$   & 3.48, 4.12, 5.26 \\ \hline
\end{tabular}
\label{tab:2}
\end{center}
\end{table}

\begin{figure}
\centering
\includegraphics[width=1.0\linewidth]{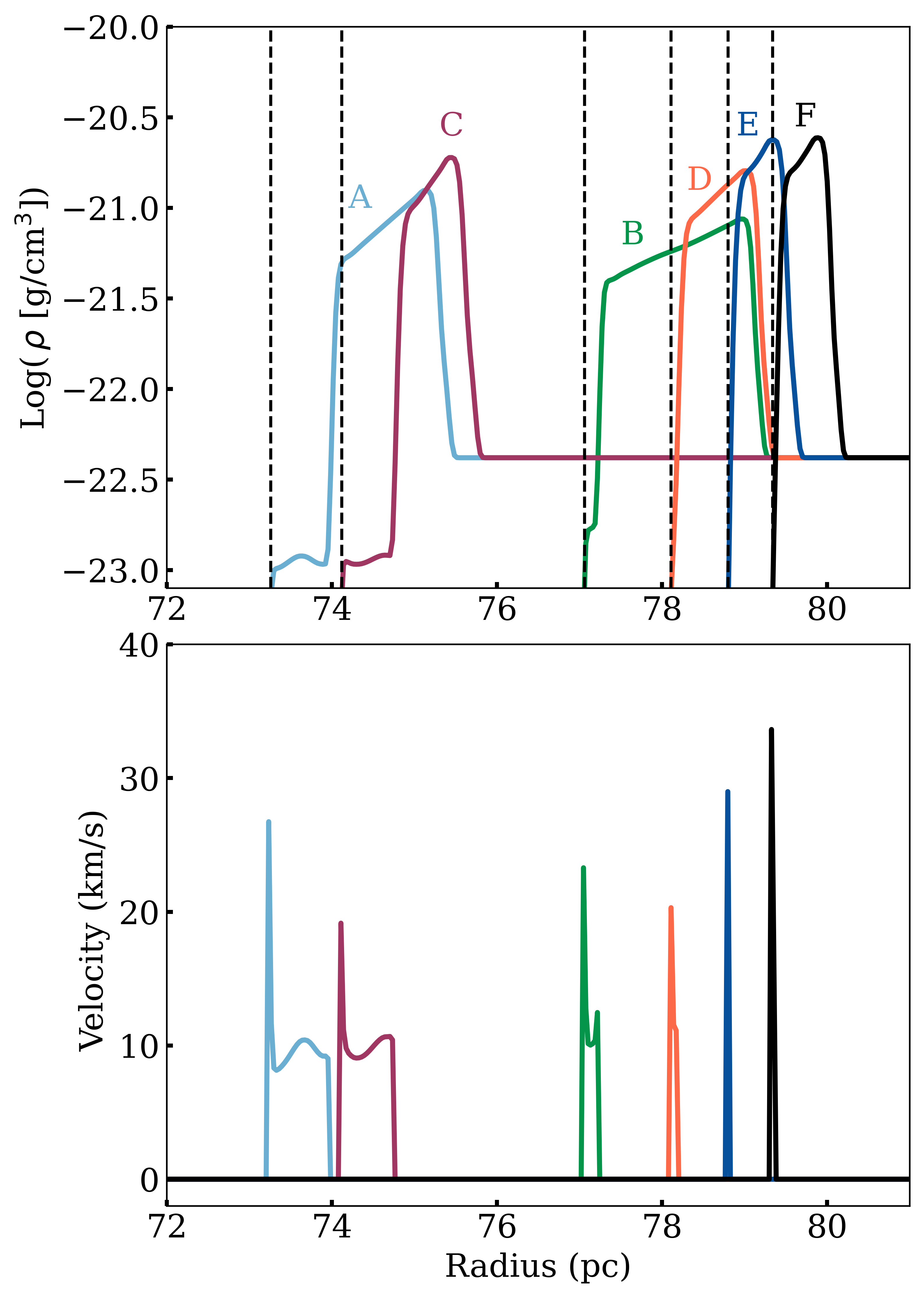}
\caption{Simulations of superbubbles with different numbers of massive stars. The top panel shows the logarithm in base 10 of mass density as a function of the r coordinate in pc. The bottom panel shows the velocity of the photoionised gas in units of km\,s$^{-1}$ as a function of the r coordinate. The letters at the top of each density curve denotes the model names from Table~\ref{tab:2}. The black dashed lines indicate the position of the contact discontinuity of each model.}
\label{fig:stars_comparison}
\end{figure}

The results of the simulations are shown in Figure~\ref{fig:stars_comparison}. This Figure presents the results of the SB for the gas density and velocity of the photionised gas in the top and bottom panels, respectively, after 5\,Myr of evolution. We show this particular age since at this stage the two most massive stars should have exploded as SNe according to DEM L50's IMF \citep{Oey1996_dynamics}.

We find that models B, D, E and F produce SBs with the largest radii ($R\lesssim$80\,pc). Models A and C produce the smallest SBs, with $R\approx$75\,pc. 
This shows that SBs can be separated into two distinct phases or groups: wind-driven and SN-driven. The wind-driven phase is temporary and occurs when no SN events have taken place within the SB. During this phase, the expansion of the SB is primarily driven by the energy from stellar winds. However, once a SN event occurs within the shell, the SB enters the SN-driven phase. At this point, the amount of energy imparted into the SB is significantly greater than that from stellar winds alone. Consequently, the SB experiences a rapid acceleration, resulting in the formation of larger SBs compared to the wind-driven phase. The transition between these two phases can be likened to an on-off switch, with the occurrence of a SN event serving as the trigger that shifts the SB into its SN-driven phase.

Figure~\ref{fig:stars_comparison} also shows that even when there are massive stars present after a SN event (as seen in model C), the resulting shock due to winds and a single SN is not sufficient to create a larger SB compared to two subsequent SNe. 
It is also interesting to note that model B, which does not have stars producing winds after the SNe explode, results in a SB that is similar to the SB for model D.

For the photoionised gas velocity (bottom panel of Figure~\ref{fig:stars_comparison}), all models produce similar velocities that are in the range of 20--35\,km\,s$^{-1}$.  
For the case were there are not any stars left (model B), the contact discontinuity is smaller as compared to model D.

These simulations result shows that separating gas into its constituent parts (e.g X-ray, photoionised, and neutral gas) rather than treating the SB as an object containing all this sub-structures, can show us the real differences between models. For the models in Figure~\ref{fig:stars_comparison}, we find that even when the complete SB formed by two subsequent SNe (model B) is similar in size to cases D--F, the contact discontinuity shows the true differences in size between these models. 
Another interesting feature is that cases D, E and F show a systematic increase in velocity that is due to the different numbers of 40\,M$_{\odot}$ stars in the model.

The latter result indicates that the radius of a SB is more heavily impacted by the number of SNe events than by the winds of the stars remaining in the cluster. In contrast, the kinematics of the SB is more dependent on the number of massive stars producing winds within the cluster.

We found that the size of the SB is sensitive to the number of SNe we include in the model, rather than including the full stellar population. This highlights the importance of knowing the original mass distribution of stars. We suggest, that Gr3 is the mechanical-energy source of DEM L50. Unfortunately, there are no studies resolving Gr3 that could let us know the distribution of massive stars in this region of DEM L50. This is why in order to create an energy input for DEM L50, we adopt the number of massive stars that are presented in \cite{Oey1996_dynamics}.
As we have shown so far, using a fraction of the massive stars, as long as we account for past SNe, should be enough to produce a model for DEM L50.

To select the combination of stars, we used the following conditions: i) The stellar IMF of DEM L50 predicts two massive stars that went SNe \citep{Oey1996_dynamics}. ii) The H-R diagram presented in \cite{Oey1996_stellarcontent} shows the presence of three 40\,M$_{\odot}$ stars in DEM L50.

We note that condition i) eliminates models A, B and C given that the present-day number of massive indicates that at least two SNe must have occurred already. Between models D--F, only model F that meets condition ii). Due to this, we choose to use the parameters of model F for the rest of the simulations.



\section{Results}\label{sec:Results}

\subsection{Overall model morphology}
\begin{figure*}
    \centering
    \includegraphics[width=1.0\linewidth]{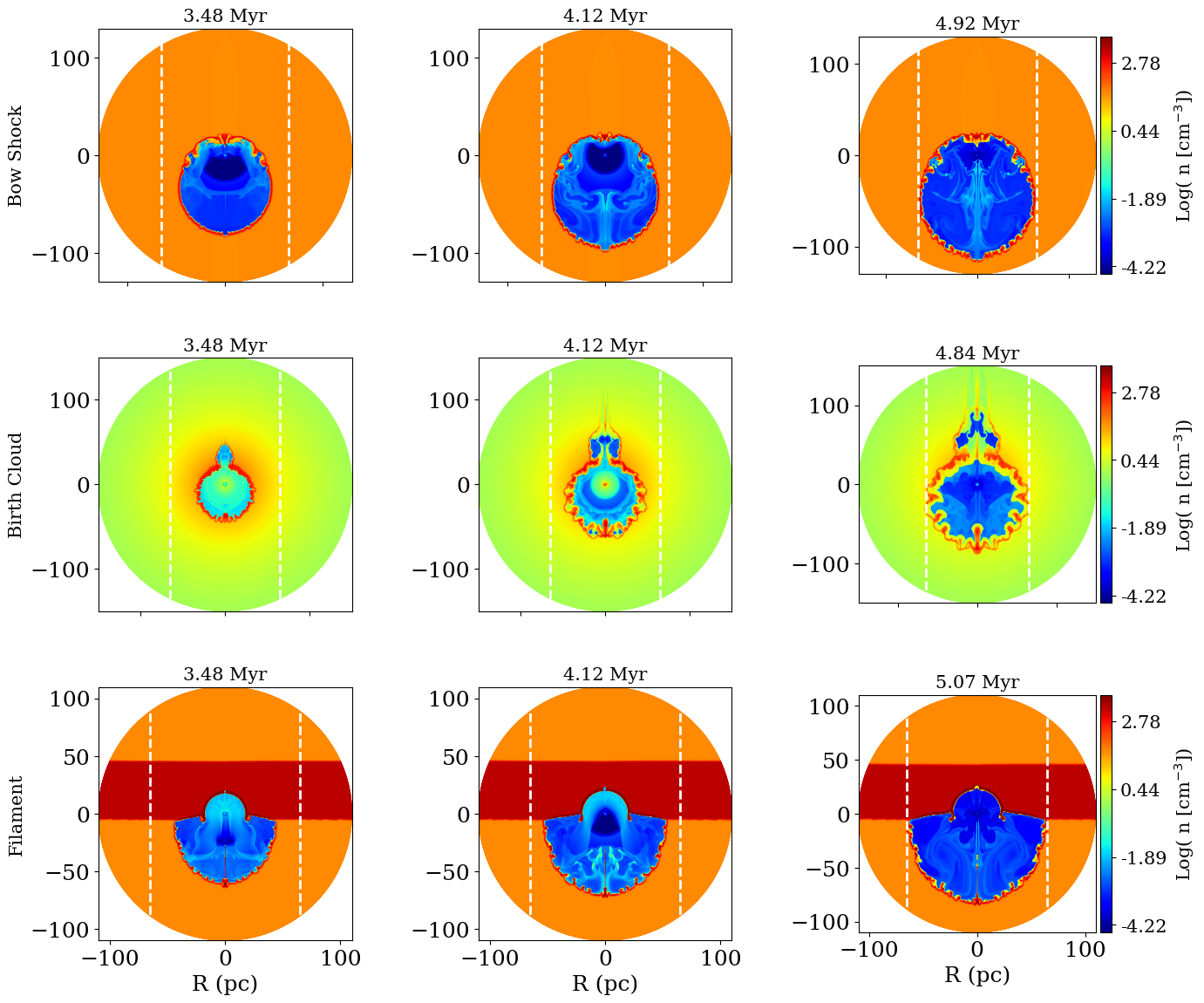}
    \caption{Number density ($n$) maps of the three different scenarios displayed at three different times. Each row shows a different scenario given by the y-axis label. The first two columns corresponds to times just after a SN explosion. The third column shows the times at which each scenario reproduces the observed diameter.
    The colour bar displays number density. The (white) dashed-line shows the measured major axis of the object, which has a true size of 130\,pc.}
    \label{fig:models}
\end{figure*}

In this section, we show the results of our models for SB DEM L50 using the initial conditions of Section~\ref{subsec:conditions}. Figure~\ref{fig:models} shows density maps for each scenario. Each row shows a different SB scenario and each column a different age. The first two columns show the morphology of the model after a SN explosion. This figure shows that none of the models reproduce the observed optical diameter of the SB, which is denoted by the white dashed lines, just after the SN explosions. The last column shows the time at which each individual model reproduces the observed optical diameter of the SB. The colour bar shows the number density ($n$) measured in cm$^{-3}$. 

All of the models produce a simple bubble-like structure, similar to the one described in the \cite{Weaver1977} model. This structure is divided as follows: a freely-expanding wind region, a shocked wind region that fills most of the volume of the SB, a contact discontinuity separating the hot shocked gas from the ionisation front, and a shell of neutral gas that separates the ionisation front from the ambient gas. 
Comparing the size of the SB models, the bow shock and the filament scenarios expand faster than the birth cloud scenario. This is due to a higher ambient gas density pressure in the birth cloud compared to the other two scenarios.
At the time the three scenarios reproduce the observed diameter, we see that the bow shock and filament scenarios produce the most asymmetric shapes. In the following sections, we will analyse in depth the individual gas structures of each SB model and we will compare them with observations of DEM L50.

\subsection{SB Components}
\begin{figure*}
    \centering
    \includegraphics[width=1.0\linewidth]{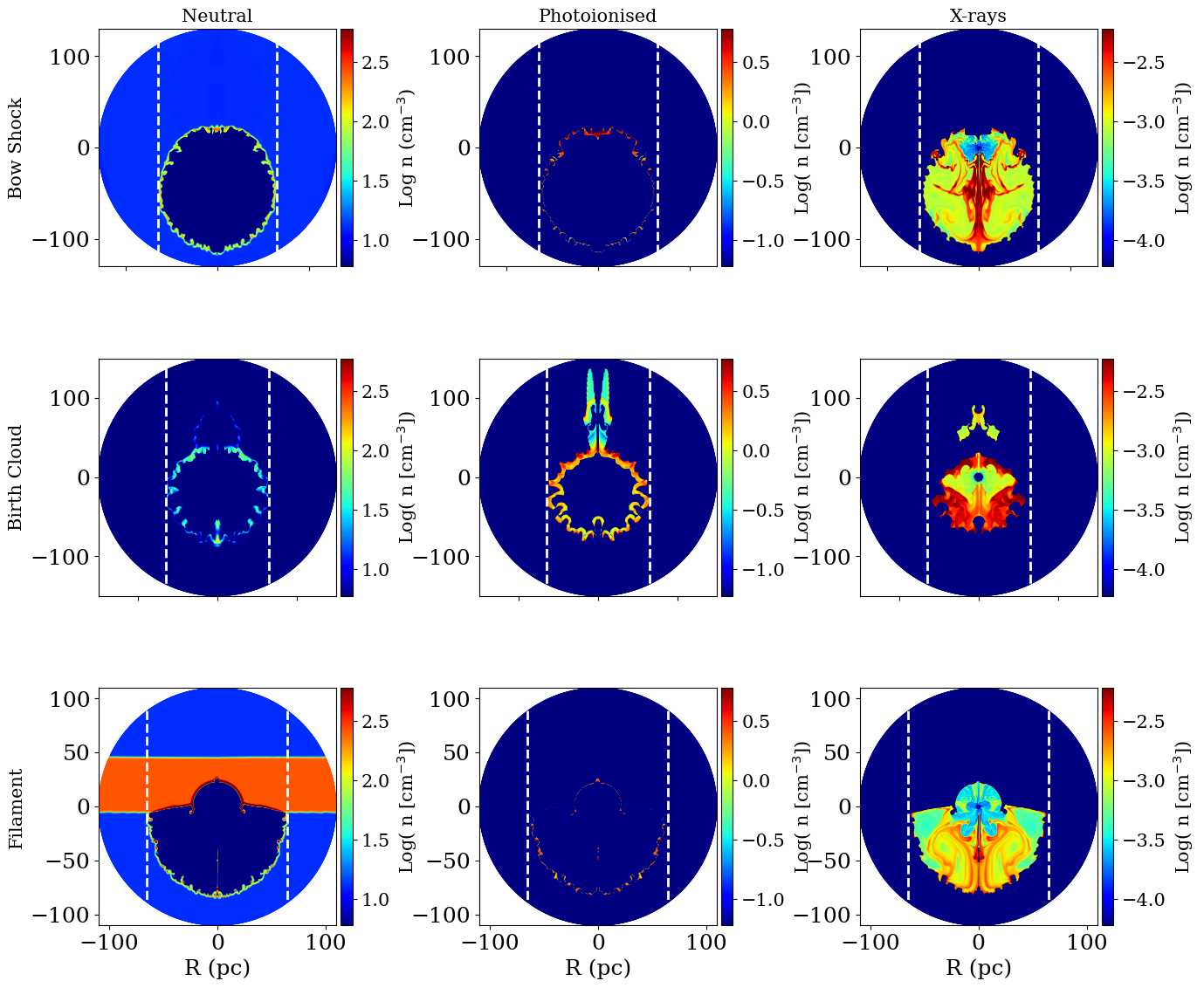}
    \caption{Density maps of three gas phases, as given by the column titles, for our models. Each panel displays the age when the model reaches the observed radius. Each row is denoted by the model name which is given by the y-axis. Each column shows: neutral gas, photoionised gas, and X-ray emitting gas. The colour bar shows the number density}
    \label{fig:densities}
\end{figure*}

In this Section, we analyse the individual gas components that make up our SB models and confront them with observations. To do this, we sectioned our simulations by gas temperature. This allowed us to produce density maps for neutral (10$^{0}$\,$\leq$\,T\,<\,10$^{4}$\,K), photoionised (10$^{4}$\,$\leq$\,T\,$\leq$\,10$^{5}$\,K), and X-ray-emitting gas (5\,$\times$\,10$^{5}$\,$\leq$\,T\,$\leq$\,10$^{9}$\,K). Figure~\ref{fig:densities} shows the resulting density maps. Here, each column presents a distinct gas characteristic, as indicated by the column title, and each row displays a different SB scenario. For each individual model we show the age when it reproduces the observed diameter according to Figure~\ref{fig:models}. As in Figure~\ref{fig:models}, the white dashed lines show the SB diameter. The colour bar shows the number density.

\subsubsection{Neutral gas}
\label{neutral_gas}


\cite{Oey2002} presented observations of the H\,{\sc i} environment in DEM L\,50. In \cite{Oey2002}'s work, they found that DEM L50 associated H\,{\sc i} is likely a neutral gas shell envelope outside of the observed optical nebula. This structure of the SB is similar to what is predicted by the \cite{Weaver1977} model on which the ambient gas and the H\,{\sc ii} region associated to the bubble is separated by a neutral gas shell. As shown in Figure~\ref{fig:densities}, our three scenarios show this same trend on which the neutral gas is surrounding the photoionised gas of the SB.
\citet{Oey2002} also calculated the swept-up mass of H\,{\sc i} of the SB. This value is shown in Table~\ref{tab:3}. 
We calculated the swept-up mass of neutral gas for our three scenarios and the results are also listed in Table~\ref{tab:3}. We find that the scenario that produces the closest value to that of DEM L50 is the filament (40\% percentage difference between model and observation), then the birth cloud (88\%), and lastly the bow shock (93\%).

\begin{table}
\begin{center}
\caption{Columns 1--3: swept up mass of neutral gas  and shell velocity for the photoionised gas calculated for our three models. Column 4: distance of the star cluster to the centre of the SB. The last row shows the measured swept up mass of H~{\sc i} obtained from the Australia Telescope Compact Array (ATCA) observations. This quantity is reported in \citet{Oey2002} and has a 32\% uncertainty. The shell velocity was obtained from H$\alpha$ observations with the 1.5 m ESO telescope using Fabry-Perot interferometry. This values are reported in \citet{Rosado1990} and have a 10-15 and 20\% uncertainty respectively.}
\begin{tabular}{ccccc}
\hline
Scenario    & Mass                    & $V_\mathrm{west}$ & $V_\mathrm{south}$ & r    \\
            & (10$^{5}$\,M$_{\odot}$) & (km\,s$^{-1}$)    & (km\,s$^{-1}$)     & (pc) \\ \hline
Bow shock   & 4.67                    & 26                & 32                 & 50   \\
Birth cloud & 0.66                    & 20                & 24                 & 17   \\
Filament    & 1.14                    & 21                & 44                 & 38   \\ \hline
DEM L50     & 1.70                    & 25                & 25                 & \dots  \\
\hline
\end{tabular}
\end{center}
\label{tab:3}
\end{table}


\subsubsection{Photoionised gas}
\label{photoionised_gas}

Now, we will compare the gas structure and morphology of photoionised gas in our models with an optical observation of DEM L50. We start by comparing the symmetry of our SB models with each other. As shown in Figure~\ref{fig:models} and \ref{fig:densities}, the bow shock and the filament scenarios result in the most elliptical SB shapes. The birth cloud scenario results in a structure that is overall spherical, even though we set an off-centre star cluster from the cloud core. 
Comparing the bow shock and the filament models, we see that both produce highly elliptical SB shapes, but the bow shock produces an asymmetric SB that has its major-axis in the opposite direction of DEM L50.

In addition to the SB morphology, we also consider the gas structures that are formed in each scenario. 
One of the most important features are the hydrodynamical instabilities, these structures help to mix hot and cold gas in the mixing layer of the SB, which will affect the total X-ray luminosity \citep{Toala2018}. 
In Figure~\ref{fig:DEML50}, we see that the ionised gas structure of DEM L50 looks like a corrugated pattern that is most noticeable at the south-west region of the object. The bow shock scenario produces two photoionised gas structures with a box-like shape to the north of the SB. When comparing the results from this scenario with DEM L50, however, there is no evidence of a similar structure in the optical shell. The birth cloud scenario produces an ionised gas pattern on which the instabilities emerge of the photoionised gas shell. This type of pattern is not observed in DEM L50, which makes the birth cloud scenario less likely. In contrast, the filament scenario produces instabilities that could be seen as corrugations in the elliptical region of the SB.

Turning our attention to the filament model, we also find that this configuration produces at first a spherical bubble inside of the filament. This bubble coincides with the position of a bubble at the northern side of DEM L50, which is powered by the stars of the star cluster Gr3 \citep{Rosado1990}. In Figure~\ref{fig:DEML50}, we highlighted the position of the bubble associated with Gr3 with a cyan circle. By making this comparison we can observe that the filament model is pretty similar to the observed SB. This makes the filament model a stronger candidate to explain the SB morphology and its properties in comparison to the other two scenarios.


\subsubsection{X-ray-emitting gas}

Lastly, we will show the resulting X-ray gas structure. The last column in Figure~\ref{fig:densities} shows the X-ray emitting gas density for the SB models, as in the the neutral and photoionised cases, this density map shows the distribution of the X-ray gas at the time when the model reproduces the observed diameter. In all the models, we see that the SB is filled with hot X-ray emitting gas that would be observed as diffuse emission in real objects. This is a behaviour that has been observed in several SBs \citep{Chu1990, Cooper2004, Jaskot2011, Rodriguez2011, Reyes-Iturbide2014, Kavanagh2015}.

We can also compare the structures that form inside the SBs. In the bow shock case 
the model has a high X-ray emitting gas density with values of $n>10^{-3.0}$~cm$^{-3}$. In these results, the highest values of density are achieved near the centre of the SB. This suggest that the gas is hottest and densest in this region compared to other parts of the SB. 
In the birth cloud scenario, the resulting X-ray density maps show that contrary to the bow shock case the density is higher around the edge of the SB and creates an envelope around the centre where the X-ray emitting gas density is lower.
Here 
the maps indicate that the X-ray emitting gas density is low compared to the previous case ($n<10^{-3.5}$~cm$^{-3}$). 
In the filament scenario, the X-ray emitting gas density maps span in a range range of values which shows that this model is an intermediate scenario, being denser that the latter model but no as much as the former. 
The difference in the range of values between models is explained by the time delay between the last SN explosion and when the model reproduces the observed diameter.
The shock from the SN raises the temperature of the surrounding gas, mixing it and temporarily boosting the X-ray-emitting gas density. This causes a shift towards higher values. As the gas cools over time, the density decreases, and the values move back towards the lower value scale.

\subsection{Gas kinematics}

\begin{figure*}
    \centering
    \includegraphics[width=1.0\linewidth]{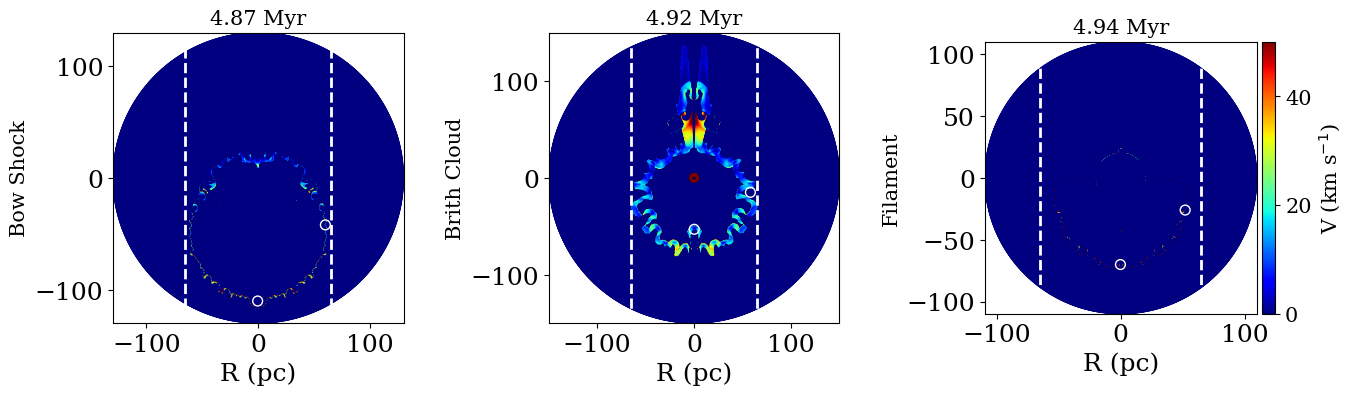}
    \caption{For the three models, maps of the speed of the ionised gas.
    The first panel shows the magnitude of the velocity for the bow shock model. The second panel shows the magnitude of the velocity for the birth cloud model. The third panel shows the magnitude of velocity for the filament model. The tittle on each panel shows the time at which each model reached the observed radius of the object. The colour bar shows the magnitude of velocity of the ionised gas.}
    \label{fig:kinematics}
\end{figure*}

We aim to show how the initial conditions for each simulation affect the velocity of the photoionised shell, and how this differences in the velocity of each simulation shaped the SB morphology. On the other hand, we aim to compare shell velocity values with the value for the optical shell in the study of DEM L50 by \citet{Rosado1990}. 
Figure~\ref{fig:kinematics} shows velocity maps of our three scenarios. As in previous sections, we show the velocity map for the time when the model reproduces the observed diameter. 
The results from our simulations can be compared with previous observational studies such as the kinematical studies of this SB using Fabry-Perot interferometry \citep{Rosado1990}. 
The bow shock model shows a systematic increase in its velocity as we move clockwise from the northern side of the SB to the southern side. In the northern section of the SB the velocity is 6\,km\,s$^{-1}$ and in the south the velocity is 32\,km\,s$^{-1}$. Here, the expansion of the SB in the northern region is being dampened by the incoming density flow due to the constant velocity we set for this model. The southern region of the SB is not affected by this density flow and the SB is able to freely expand into the ambient space. This difference in velocity is what help the formation of the elliptical shape of this particular SB model. 
Overall, the birth cloud model shows a uniform velocity distribution. For this model, the velocity at the northern side is 17\,km\,s$^{-1}$ and 26\,km\,s$^{-1}$ at the southern region. As shown in Figure~\ref{fig:kinematics}, in this case the variations in the velocity in the entire shell are not high in comparison to those from the bow shock model. The difference in the velocity from north to south in this case is explained by the ambient density profile. As the SB starts to expand at the beginning of the simulation, it encounters the dense cloud core (n = 10$^{4}$ cm$^{-3}$), which slows down the SB expansion in the north portion of the shell. 
In the south, the SB expands into an environment with a decreasing ambient density, resulting in higher velocities. 
The filament model shows a similar behaviour as that of the bow shock case. In the filament case, the north velocity is 26\,km\,s$^{-1}$ and the south velocity is 39\,km\,s$^{-1}$, here the difference between the north and south velocity is not as big as in the bow shock model. 
The difference in velocity is due to the different densities in the environment. In the north side, the SB has to grow in a high-pressure environment due to the filament density. In the south region, the SB grows in an environment with a lower pressure, thus the gas is able to achieve higher velocity values. \\
Let us compare the velocity of the photoionised shell with that reported for DEM L50 (last row of Table~\ref{tab:3}). For this comparison, we report the velocity values in two regions of our models that are shown with empty dashed-circles in each panel of Figure~\ref{fig:kinematics}. This values are also shown in Table~\ref{tab:3}. We find that all the models can reproduce the observed velocity within the uncertainty measure.

\subsection{Off-centre SNe and X-ray luminosity} 
\label{luminosity_X}
In \cite{Chu1990}, the authors presented observations of diffuse X-ray emission for SBs in the LMC. One of the main results of that work is that some SBs can have high luminosities, which cannot be explained using the \citet{Weaver1977} model. In particular, DEM L50 is also a SB with a high X-ray luminosity value as also showed more recently by \citet{Jaskot2011}. In \cite{Chu1990}, a solution offered to reconcile models and observations is to account for off-centre SNe explosions inside the SB. 
Recent works have modelled SBs using the spatial distribution of the stars in the cluster, aiming to understand if this could explain the observed soft X-ray luminosity. In particular, see the study of N\,70 by \citet{Schneiter2022}. Although, the model of the latter authors can explain the observed properties of N70, it cannot reproduce the soft X-ray luminosity without introducing an artificially off-centre SN. 
We have shown with our results that by taking into consideration a more complex environment for the ISM, we can in fact produce off-centre SNe naturally. 
In Table~\ref{tab:3} we show the distance between the region where winds and SNe are being produced in our simulations and the geometrical centre of the SB. The distance has been measured at the age shown in the last column of Figure~\ref{fig:models}. 
We have selected this age to determine the farthest distance from the SB centre that we can achieve with our models. We found that for every model we get distances larger than 10\,pc. 
To estimate the soft X-ray luminosity, we integrate the emission following \cite{Chu1990}:

\begin{equation} \label{eq:6}
    L_\mathrm{X} = \int n(r)^{2}\,\Lambda[T(r)] d^3\,r.
\end{equation}
\noindent Here, $n$ is the density measured from the X-ray observations in cm$^{-3}$ and $\Lambda[T(r)]$ is the emissivity of hot gas. According to \citet{Garcia-Segura1995}, we can estimate the luminosity by assuming a constant $\Lambda$, with a value $\Lambda$ = 9$\times$\,10$^{-24}$ erg\,cm$^{-3}$\,s$^{-1}$ for $T\ge5\times10^{5}$~K. Originally, this emissivity was used to model the luminosity observed with the \emph{ROSAT} X-ray observatory. By using the above method, we need to be aware that the values obtained are low-limit values \citep{Garcia-Segura1995}. 
Although, this approach could be considered old-fashioned, the obtained values will give us an idea if our models can in fact produce an X-ray luminosity that is in the right ballpark. This approach will enable us to compare the luminosities that are reported in \cite{Jaskot2011} for different emission models of DEM L50, such as single and two-temperature or non-equilibrium ionisation models (NEI).

\begin{figure}
    \centering
    \includegraphics[width=1\columnwidth]{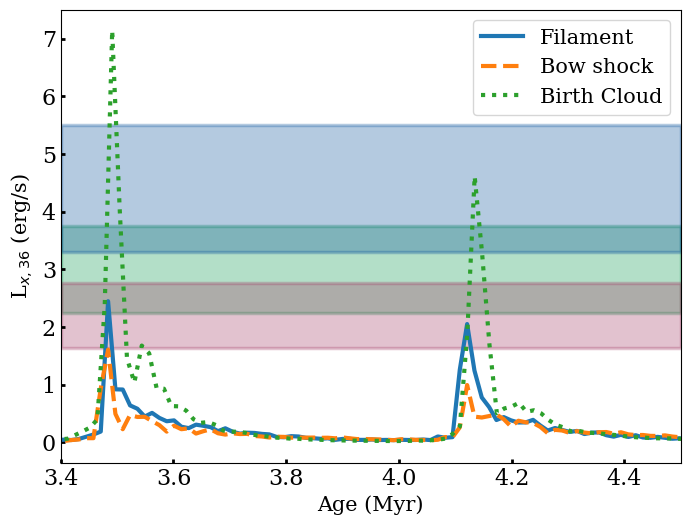}
    \caption{Temporal evolution of X-ray luminosity for our three scenarios. The green filled with squares shows the estimated luminosity value of DEM L50 in the 0.3--2.0 keV energy band of {\it Chandra} for a single temperature spectral fit. The blue area filled with slashes shows the calculated luminosity from a two temperature spectral fit. The pink area filled with circles show the calculated luminosity for a plasma in non-equilibrium ionisation. All the luminosities include the uncertainty of the fit. All the luminosities have been taken from \citet{Jaskot2011}.}
    \label{fig:luminosity}
\end{figure}

Figure~\ref{fig:luminosity} shows the luminosity for our three scenarios as a function of time in Myr, the three coloured regions show the estimated luminosity with the plasma emission models for DEM L50 as reported in \cite{Jaskot2011}.
The green, blue, and pink areas correspond to single, two-temperature, and NEI plasma emission models for DEM L50.
A gas in NEI indicates that due to recent heating by a shock (such as one produce by a SN), there is no balance between collisional ionisation from the ground states of the various atoms and ions in the gas and the process of recombination from the higher ionisation states \citep{Dopita2003}. 

The resulting X-ray luminosities, shown in Figure~\ref{fig:luminosity}, have two maximum values which correspond to two SN explosions caused by the two most massive stars in our simulations. During the fist SN explosion, the model that produces the highest value of $L_\mathrm{X}$ is the birth cloud scenario ($L_\mathrm{X}\approx7.5\times10^{36}$~erg\,s$^{-1}$).
This high value is due to the size of the SB at the time of the SN. In comparison to the other two models, the SB is very small, as shown in the first column of Figure~\ref{fig:models}. 
Because of this, the shock generated by the SN reaches the shell before becoming subsonic, causing the high luminosity value. 
According to \cite{Oey1996_dynamics}, the observed soft X-ray luminosity of DEM L50 is attributed to a 60\,M$_{\odot}$ star that went SN at 4.12\,Myr.
Hence, we are interested in the second maximum of Figure~\ref{fig:luminosity} given that this is caused by the 60\,M$_{\odot}$ star that exploded as a SN. At this time, the model that produces the highest luminosity is the birth cloud scenario ($L_\mathrm{X}=4.6\times10^{36}$~erg\,s$^{-1}$), followed by the filament ($L_\mathrm{X}=2.1\times10^{36}$~erg\,s$^{-1}$) and by the bow shock ($L_\mathrm{X}=1.0\times10^{36}$~erg\,s$^{-1}$). 
We find that all three models predict a high luminosity, but only the filament is in the expected range for DEM L50, which agrees with a plasma model that is in NEI. This is coherent with the idea of recent heating due SN activity, and explains the non-equilibrium state of the gas. 

\begin{figure*}
    \centering
    \includegraphics[width=1\linewidth]{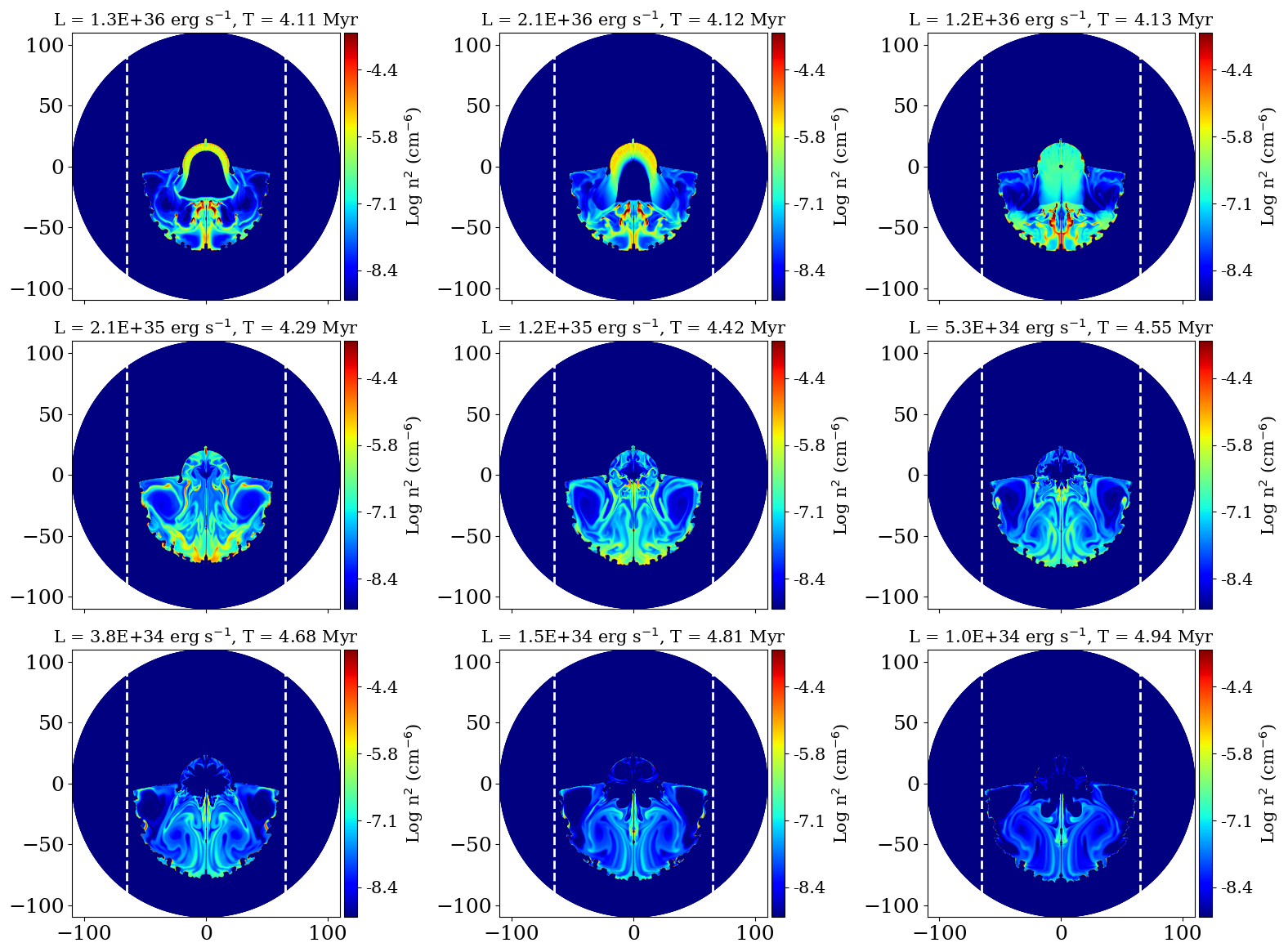}
    \caption{Density maps of the X-ray-emitting gas (5$\times$10$^{5}\leq T\leq10^{9}$ K). Each panel represents a time snapshot of the filament scenario. Each panel has its total estimated X-ray luminosity in erg\,s$^{-1}$ units and its corresponding time in Myr. The colour bar shows the density in units of cm$^{-3}$.}
    \label{fig:xray_timelapse}
\end{figure*}

Our results indicate that the filament model produces the closest X-ray luminosity value among the three scenarios we examined, and makes it our best model because it has the best overall agreement with other observable quantities and the closest morphology to DEM L50.
Due to the filament model being our best scenario, we are also interested in comparing the X-ray gas structure of this model with the X-ray imaging of DEM L50. For this, we produced the emission maps shown in Figure~\ref{fig:xray_timelapse}. This Figure shows the temporal evolution of the X-ray gas emissivity in the SB model after the second SN. The title of each panel shows the estimated X-ray luminosity for a given age in Myr. 
We find that the SB is X-ray bright in some regions of the SB and the structure immediately after the SN is rich in filaments. 
Between $t$ = 4.13 and 4.29\,Myr, the structure of the gas is similar to that of DEM L50 (Figure~\ref{fig:DEML50}), specially at the southern region of the X-ray shell.
At $t$ = 4.29\,Myr, the bright cyan filaments are very similar to those observed in DEM L50 for the 0.6--1.6~keV energy range (figure 4b in \citealt{Jaskot2011}).
As shown in Figure~\ref{fig:xray_timelapse}, our models do not match the observed luminosity at the age when they reproduce the dimensions of the SB. This 
highlights the limitations of our current understanding of the environment and stars that formed the SB. In addition, stellar evolution models only provide us with a limited set of masses, which we use to infer the number of stars in the IMF. Therefore, there is also some degree of uncertainty about the time at which the SN event occurred, as the massive 60\,M$_{\odot}$ star could also be a 55\,M$_{\odot}$ star that went SN at a later time. 
As we mentioned before, using Eq.~\ref{eq:6} could be an outdated procedure for calculating the true soft X-ray luminosity of our models. A more up to-date way of performing this calculation is by modelling a synthetic X-ray spectrum using the density and temperature of our simulations in conjunction with a set of abundances for the ISM. Using this information and the software package {\sc ChiantiPy} \citep{Dere2019}, we can calculate an X-ray spectrum for our models. Then, we integrate this spectrum in the energy range of interest, for instance, the soft X-ray range of {\it Chandra}. 
We calculated the luminosity using {\sc ChiantiPy} for the time bin with the highest luminosity in the filament model. 
To do this, we adopted the LMC's ISM abundances of \cite{Russell1992}, and we integrated the resulting spectrum in the 0.3--2.0~keV energy range, which is the energy range for soft X-rays in {\it Chandra}. 
We obtain a value of $L_\mathrm{X}$ = 5.8$\times$10$^{34}$\,erg\,s$^{-1}$, which in comparison with the value of $L_\mathrm{X}$ obtained by using Eq.~\ref{eq:6} is a factor $\sim$\,36 smaller. 

\begin{table}
\begin{center}
\setlength{\columnwidth}{0.1\columnwidth}
\setlength{\tabcolsep}{0.6\tabcolsep}
\caption{Summary of the properties that our models can reproduce when compared to the observed properties of DEM L50.}
\begin{tabular}{cccccc} 
\hline
Scenario  & Asymmetric & Radius     & Velocity   & Swept-up mass & $L_\mathrm{X}$ \\ \hline
Bow shock   & \checkmark& \checkmark & \checkmark & \checkmark    & \xmark                \\
Birth cloud & \xmark & \checkmark & \checkmark & \xmark           & \checkmark       \\
Filament    & \checkmark &\checkmark & \checkmark & \checkmark    & \checkmark   \\
\hline
\end{tabular}
\end{center}
\label{tab:4}
\end{table}

\section{Discussion}
\label{sec:Discussion}

We developed three models tailored to SB DEM L50 aiming to explain the high soft X-ray luminosity of this SB. This work improved previous results presented in \cite{Oey2004} based on 1D models, by testing the effects of using different conditions for the environment were the SB grows and massive star combinations. By adjusting the initial conditions, we are able to produce off-centre SNe naturally, which increases the soft X-ray luminosity produced by the models. 
In Table~\ref{tab:4}, we summarise different properties that our models can reproduce given the initial conditions. In general, the model that produces the closest properties to the observed ones is the filament scenario.
We find that by assuming a central star cluster at the edge of a filament, the resulting SB model predicts a hot bubble that is in a similar position as the one produced by the star cluster Gr3 \citep{Rosado1990}. Star formation taking place in filaments has been observed in several regions of space, such as in the TaurusB211/B212 filament \citep[e.g.,][]{Palmerim2013}. In the LMC, star formation in molecular filaments has been extensively studied in regions such as N159-W and N159-E \citep{Fukui2015, Fukui2019}. 
Furthermore, the interaction between a SB and a giant molecular filament in the Milky Way has been reported in \cite{Clarke2022}, showing that this kind of scenario is plausible.
Of course, we still need to confirm the presence of a filament near DEM L50 in order to make our filament model likely.

\begin{figure*}
    \centering
    \includegraphics[width=0.9\linewidth]{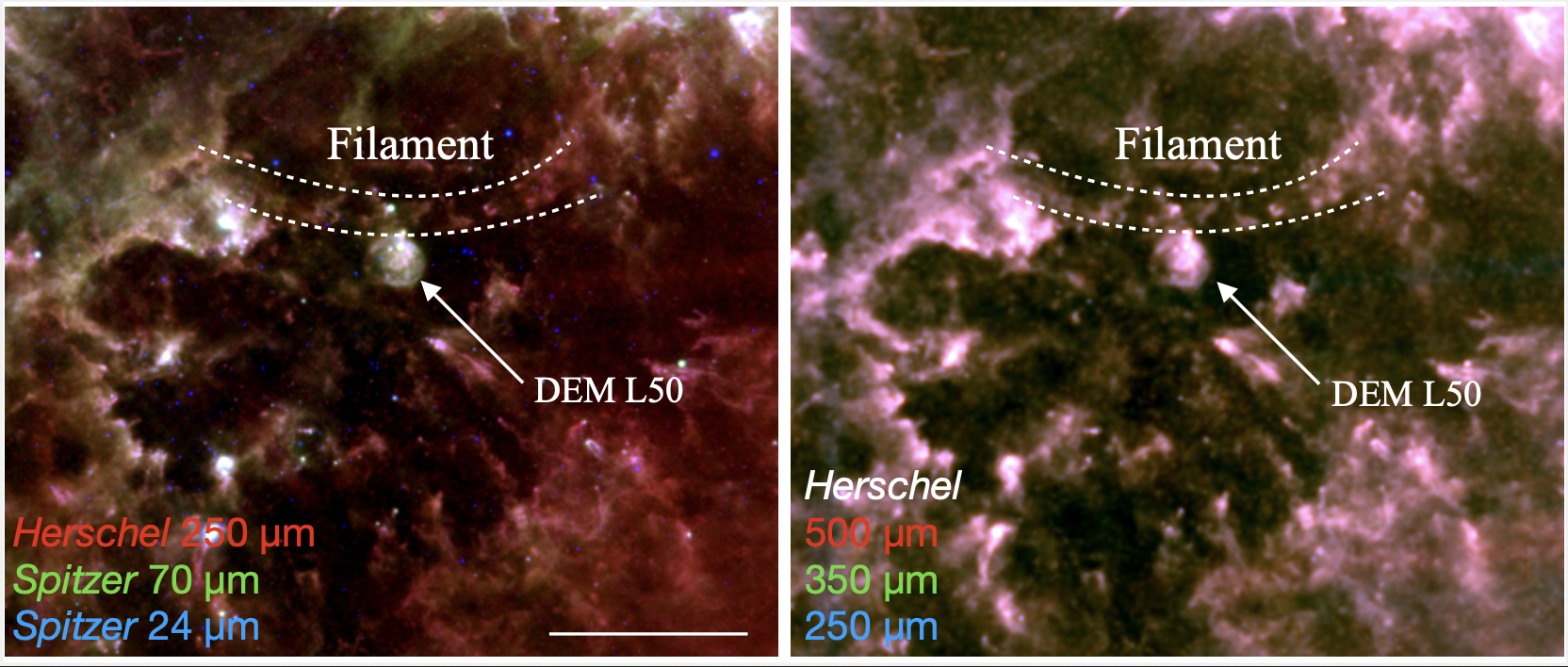}
    \caption{Colour-composite IR images of the vicinity of DEM L50. The possible filament is shown between dashed-line curves. The white segment represents 30~arcmin.}
    \label{fig:DEML50_dust}
\end{figure*}

The emission maps presented in \citet{Oey2002} show some evidence of H\,{\sc i} that is at the northern part of DEM L50. This H\,{\sc i} structure could be part of a molecular filament on which cluster Gr3 formed before the formation of the observed SB. 
Although, an HI view can give us some hint about the environment surrounding the SB, we still need more information on the extended emission of the dust component. Figure~\ref{fig:DEML50_dust} shows an IR image of the vicinity of DEM L50 obtained by combining publicly available data from {\it Herschel} and {\it Spitzer}. This image shows a dense region in the northern area of DEM L50, which has been highlighted between two dashed lines. The spatial correspondence between our numerical simulation and the IR observations shows that this could be the filament we are looking for. Thus, the filament model is likely. Of course, we need to be aware that we are looking at a 2D projection of what is in reality a 3D gas distribution but, as stated by \cite{Oey2002}, the H\,{\sc i} in this region could be an envelope that its partially interacting with DEM L50 and the dust could be in the same state.

\begin{figure}
    \centering
    \includegraphics[width=1.0\linewidth]{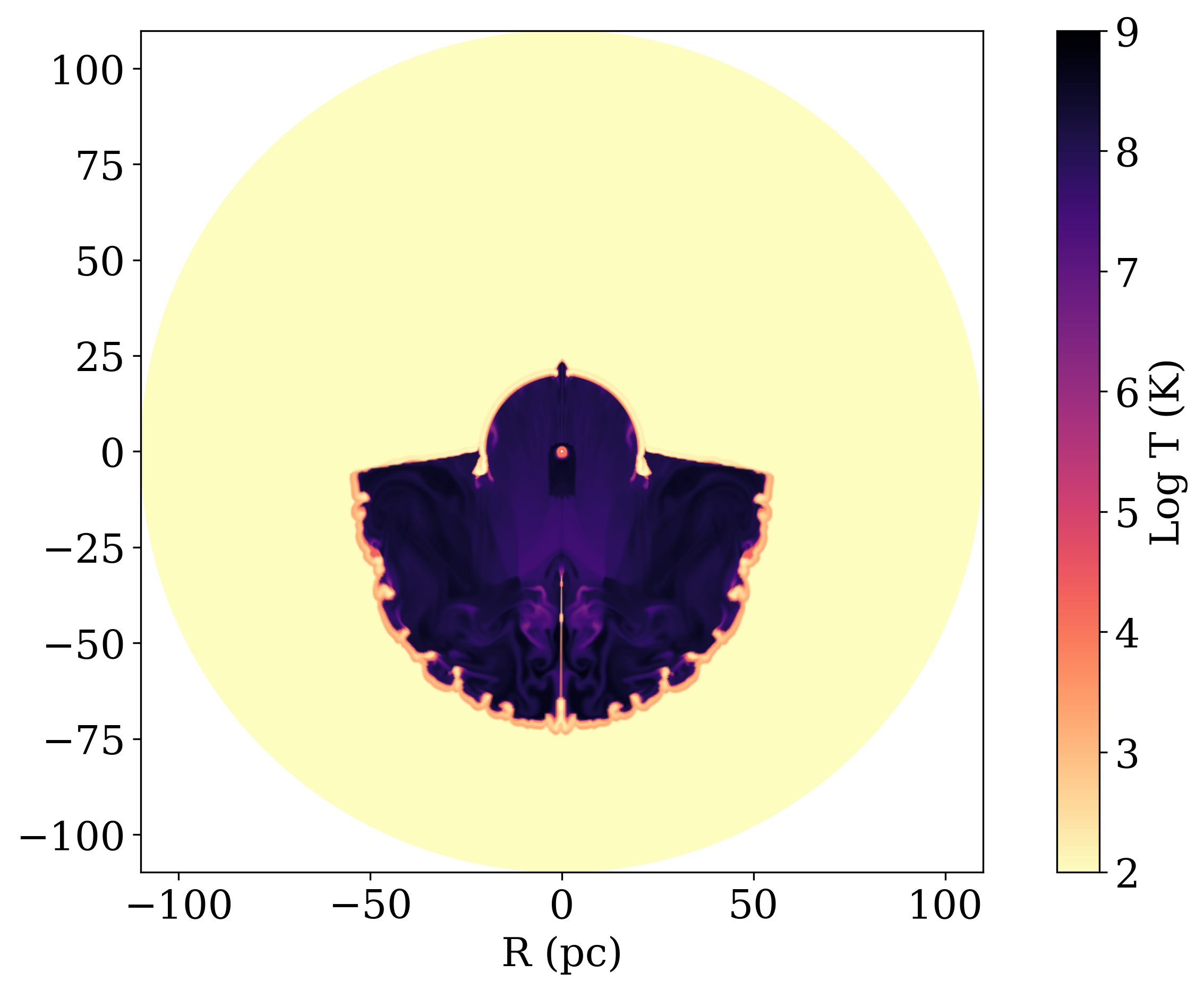}
    \caption{Temperature map for a snaphot after the second SN explosion in our simulation, at T = 4.12\,Myr.}
    \label{fig:temperature}
\end{figure}

The results presented in this paper show that the characteristics of the environment in which massive stars evolve is crucial for the subsequent evolution of their SBs. The scenarios studied here share similar characteristics as the off-centre SN explosion scenario. Assuming a constant density ambient to model a SB, it is a good first-order approximation but, it does not hold in all cases. 
Regarding the discrepancy between the values of L$_{x}$ from equation~\ref{eq:6} and {\sc ChiantiPy}. 
The luminosities from equation~\ref{eq:6} are larger than the ones from {\texttt{ChiantiPy}} by a factor $\sim$ 36 for $E$ = [0.3--2.0]\,keV. 
We think that this discrepancy is due to our models producing more soft X-ray emitting gas at temperatures between 10$^{8}$-10$^{9}$\,K.
In Figure~\ref{fig:temperature}, we show a temperature map of the filament model for the time slice corresponding to the maximum luminosity value (t = 4.12\,Myr). This plot shows that in general the SB has temperatures T > 10$^{8}$\,K. 
As showed in \cite{Toala2018}, the emissivity of soft X-rays in hot diffuse nebulae has a two-temperature component: one component with a very steep emissitivy for T = 1-3 $\times$ 10$^{6}$\,K and a second component which is more like a broad plateau for T = 10$^{7}$-10$^{8}$\,K. 
For temperatures above 10$^{8}$\,K the emissivity starts to diminish according to \cite{Toala2018}. This explains why our models do not produce a sufficiently high luminosity between 0.3--2.0\,keV. 
Using equation~\ref{eq:6} we get a higher luminosity, as we are assuming a constant value for the emissivity and thus, each temperature bin contributes the same amount to the total value.

If our models produce a high temperature, we need to decrease the temperature of soft X-rays without changing the overall dynamics of the SB. An effect that our models are not taking into account is thermal conduction. 
This diffusive process should have the effect of lowering the temperature below 10$^{8}$\,K as showed by \cite{El-Badry2019}, without changing the overall dynamics of the gas \citep{Toala2011, El-Badry2019}. 
Further studies to understand the effect that thermal conduction has in the overall spectrum and in the plasma properties (e.g luminosity and temperature) in SBs growing in non-uniform environments would be of interest due to the fact that they will provide us with a better understating of the role that this physical process plays in the evolution of the energy budget in SBs.
In a follow-up paper, we intend to study the effect that thermal conduction has in our SB models. Specifically, we will show how the X-ray properties of the filament scenario change if we include this new ingredient in our models. 
It is also worth to notice that another way of increasing the soft X-ray luminosity could be done by increasing the resolution in our models. By doing this, we could increase the instabilities formed as the SB expands. This could enhance the mixing between hot and cold material, which in turn could increase the soft X-ray luminosity in the SB. This resolution study is intended for the follow-up article on which we will compare also with the effects of adding thermal conduction.

\section{Conclusions}
\label{sec:Conclusions}

In this work, we have tested different initial conditions to produce hydrodynamical models that can explain the soft X-ray luminosity in SB DEM L50 as well as other properties. The results of this paper show that  the morphology of DEM L50 can be better explained by a configuration in which the star cluster formed at the edge of a filament. 
In our simulations, off-centre SNe, which are a mechanism to enhance soft X-rays, are produced naturally by the initial conditions of the ambient where the SB grows. 
One of the main findings of our work is that it highlights the crucial role that the environment plays in producing the observed properties in SBs. It also shows that a constant ISM could be an over simplification of the true complexity of space.\\
The best model of this work also predicts a star cluster that is in the same position as star cluster Gr3. This suggest the possibility that SB expansion is driven by small stellar associations rather than stars dispersed throughout the SB. 
An in-depth study of Gr3 and similar star clusters in other SBs could help us understand if in reality SBs are formed by stars in this type of star clusters.
When calculating the X-ray luminosity produced by our models, we found that all the models can produce high values of L$_{x}$ by assuming a constant emissivity, which are consistent with what is expected from off-centre SNe and agree with the reported luminosity of DEM L50 for a plasma in NEI \citep{Jaskot2011}.  
We showed that when comparing the luminosity obtained by assuming a constant emissivity and that obtained by integrating a synthetic spectrum over the range E = [0.3-2.0]\,keV, the later method produces a soft X-ray luminosity value that is a factor $\sim$\,36 smaller than the value we get using a constant emissivity. 
This is because our hydrodynamical models produce most of the X-ray gas at temperatures above 10$^{8}$\,K and above this threshold, the contribution of the emissivity to soft X-rays starts to decrease as stated in \cite{Toala2018}.
Including thermal conduction in our models could be a solution to this discrepancy given that this should lower the temperature of the gas by almost an order of magnitude according to \cite{El-Badry2019}. 
We intend to improve our analysis in the future by including thermal conduction in our models.\\

\section*{Acknowledgements}
A. Wofford and R. Orozco-Duarte acknowledge financial support from grant UNAM PAPIIT IN106922 (PI Wofford).
We thank Michael L.\ Norman and the Laboratory for Computational
Astrophysics for the use of ZEUS-3D. The computations
were performed at the Instituto de Astronom\'{\i}a-UNAM at Ensenada. We thank the referee for the very useful suggestions that significantly improved the manuscript.
\section*{Data Availability}
The data underlying this article will be shared on reasonable request to the corresponding author.



\bibliographystyle{mnras}
\bibliography{references} 





\bsp	
\label{lastpage}
\end{document}